\documentstyle[twoside,fleqn,espcrc2,epsfig]{article}

\newcommand{\be}{\begin{equation}}                                              
\newcommand{\ee}{\end{equation}}                                                
\newcommand{\half}{\frac{1}{2}}

\title{Numerical simulation of dynamical gluinos: experience with
       a multi-bosonic algorithm and first results}

\author{R. Kirchner \address{Deutsches Elektronen Synchrotron, DESY,
                             Notkestr. 85, D-22603 Hamburg, Germany},
	S. Luckmann\thanks{Poster presented by S. Luckmann.}
           \address{Institut f\"ur Theoretische Physik I,
                    Universit\"at M\"unster, Wilhelm-Klemm-Str. 9, \\
                    D-48149 M\"unster, Germany},
	I. Montvay\ $\rm^a$,
	K. Spanderen\ $\rm^b$,
        J. Westphalen\thanks{Talk given by J. Westphalen.}\ $\rm^a$
        \\[0.5em]
        DESY-M\"unster Collaboration \\[0.5em]}
       
\begin{document}

\begin{abstract}
 We report on our experience with the two-step multi-bosonic algorithm
 in a large scale Monte Carlo simulation of the SU(2) Yang-Mills theory
 with dynamical gluinos.
 First results are described on the low lying spectrum of bound states,
 the string tension and the gluino condensate.
\end{abstract}

\maketitle

\section{INTRODUCTION}\label{sec1}
 An important step towards the understanding of non-perturbative
 properties of supersymmetric (SUSY) gauge theories is the numerical
 simulation of the supersymmetric extension of Yang-Mills (SYM) theory.
 The theoretical motivation for this has been summarized at the
 previous \cite{EDINBURGH} and present \cite{STRASSLER} lattice
 conferences.
 (For references see these reviews and \cite{BUCKOW}.)
 First steps towards a numerical Monte Carlo simulation with dynamical
 gluinos have already been presented a year ago \cite{DESYMUNSTER}.
 Since then our collaboration made important progress in a first large
 scale simulation of SU(2) SYM on the CRAY T3E-512 at HLRZ J\"ulich.
 This is a short status report which will be followed soon by a more
 detailed publication.

\subsection{Lattice action}\label{sec1.1}
 The SU(2) Yang-Mills theory with gluinos (= Majorana fermions in the
 triplet representation) becomes supersymmetric in the massless limit.
 Massive gluinos break SUSY softly.
 For a lattice regularization one can take the Wilson action for gauge
 field and gluinos, as proposed some time ago by Curci and Veneziano
 \cite{CURVEN}.
 This contains two bare parameters: $\beta$ for the gauge coupling
 and $K$ for the hopping parameter (bare gluino mass).
 The Majorana nature of the gluino is taken into account by
 considering $N_f=\half$ adjoint Dirac flavours.
 The effective action for the gauge field assumes the form
\be\label{eq01}
S_{CV} = \beta\sum_{pl} \left( 1-\half{\rm Tr\,}U_{pl} \right)
- \half\log\det Q[U] \ ,
\ee
 where the fermion matrix is
\be\label{eq02}
Q_{yv,xu} = \delta_{yx}\delta_{vu} - 
K\sum_{\mu=\pm} \delta_{y,x+\hat{\mu}}(1+\gamma_\mu) V_{vu,x\mu} 
\ee
 with the gauge link in the adjoint representation $V_{vu,x\mu}=\half
 {\rm\,Tr\,}(U_{x\mu}^\dagger \tau_v U_{x\mu} \tau_u )$.

 An interesting new development is that an exactly zero mass gaugino can
 be described on the lattice by the {\em Neuberger-action}
 \cite{NEUBERGER}.
 (Concerning chiral symmetry see \cite{NIEDERMAYER}.)
 The technical difficulty is to determine the necessary inverse
 square-root $(Q^\dagger Q)^{-\half}$.
 For this we can use the quadratically optimized polynomials discussed
 in sec.~\ref{sec2}.
 The results of a numerical study of the Neuberger-action will be
 published elsewhere.

\subsection{Pfaffians}\label{sec1.2}
 The Curci-Veneziano action assumes $\det(Q)^\half$ for the Majorana
 fermion.
 This may lead to a sign problem because the path integral for Majorana
 fermions gives the Pfaffian
\begin{eqnarray} \nonumber
& & {\rm Pf}(M) \equiv
\int [d\phi] e^{-\half\phi_\alpha M_{\alpha\beta} \phi_\beta}
\\
\label{eq03}
& = & \frac{1}{N!2^N} \epsilon_{\alpha_1\beta_1 \ldots \alpha_N\beta_N}
M_{\alpha_1\beta_1} \ldots M_{\alpha_N\beta_N} \ ,
\end{eqnarray}
 where $M \equiv CQ$ satisfies
\be\label{eq04}
{\rm Pf}(M)^2 = \det(M) = \det(Q) = \det(\tilde{Q}) \ .
\ee
 Here $\tilde{Q} \equiv \gamma_5 Q$ is the hermitean fermion matrix.
 Since $\tilde{Q}$ has doubly degenerate real eigenvalues, the
 determinant is non-negative: $\det(Q) \geq 0$.
 The sign of the Pfaffian is, however, unknown.

 A {\em numerical procedure} for the computation of Pfaffians can be
 based on the decomposition \cite{BOURBAKI}
\be\label{eq05}
M = P^T J P ,\hspace{2em}  {\rm Pf}(M)=\det(P) \ ,
\ee
 where $J$ is a block-diagonal matrix containing on the diagonal
 $2\otimes2$ blocks equal to $\epsilon=i\sigma_2$.
 This form of $M$ can be achieved by a procedure analogous to the
 Gram-Schmidt orthogonalization and then $P$ turns out to be a
 triangular matrix.
 (See e.g.~the treatment of simplectic groups in \cite{HAMERM}.)

\begin{figure}[th]
\vspace*{-1.2cm}
\begin{center}
\epsfig{file=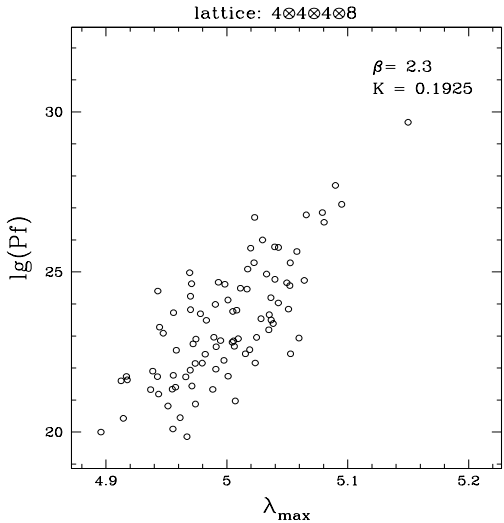,
        width=7.5cm,height=7.0cm,
        angle=0}
\vspace*{-1.4cm}
\caption{\label{fig01}
 The values of the pfaffian on $4^3\cdot8$ lattice at
 $(\beta=2.3,K=0.1925)$ versus the largest eigenvalue $\lambda_{max}$
 of $Q^\dagger Q$.}
\end{center}
\vspace*{-1.1cm}
\end{figure}

 The numerical procedure requires the storage of a full $N \otimes N$
 matrix, similarly to the determinant calculation by LU-decomposition.
 Therefore, one can only deal with relatively small lattices.
 In a test with dynamical gluino updating at $(\beta=2.3,K=0.1925)$ on
 $4^3\cdot8$ lattice the Pfaffian of every randomly chosen configuration
 turned out to be positive (see fig.~\ref{fig01}).

\begin{figure}[th]
\vspace*{-1.2cm}
\begin{center}
\epsfig{file=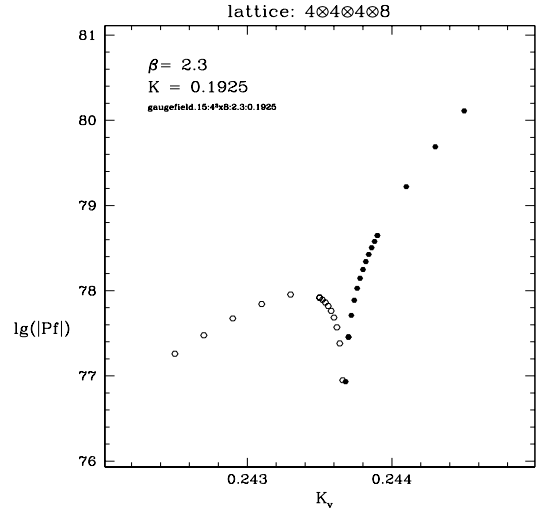,
        width=7.5cm,height=7.0cm,
        angle=0}
\vspace*{-1.7cm}
\caption{\label{fig02}
 The absolute value of the pfaffian on a $4^3\cdot8$ configuration as a
 function of the hopping parameter in $M$ (and $\tilde{Q}$).
 Open points stand for ${\rm Pf}(M)>0$, full ones for ${\rm Pf}(M)<0$.}
\end{center}
\vspace*{-1.1cm}
\end{figure}

 In general, however, the Pfaffian is not always positive.
 In fact, if a pair of degenerate eigenvalues of the hermitean fermion
 matrix $\tilde{Q}$ changes sign, it is plausible that the sign of
 ${\rm Pf}(M)$ changes, too.
 An example is shown by fig.~\ref{fig02}, where one of the configurations
 in fig.~\ref{fig01} is considered as a function of the ``valence''
 hopping parameter $K_v$ in $M$.
 As the numerical determination of the nearly zero eigenvalues of
 $\tilde{Q}$ shows, at the same $K_v$ where ${\rm Pf}(M)$ changes sign
 there is also a sign change of an eigenvalue pair.

 The spectral flow of $\tilde{Q}$ is relevant for the
 {\em overlap-inspired fermionic definition} of the topological charge
 (see \cite{EHN_FLOW} and references therein).
 Since the index of a massless Dirac operator in the adjoint
 representation of SU(2) in a gauge field background of topological
 charge $Q_{top}$ is $4Q_{top}$ \cite{INDEX}, this suggests that at
 the hopping parameter value $K^\prime$ used in the Neuberger action
 the sign of ${\rm Pf}(M)$ is given by
\be\label{eq06}
{\rm Pf}(M(K^\prime))/|{\rm Pf}(M(K^\prime))| = e^{2\pi i Q_{top}} . \
\ee
 One has to have in mind that the fermionic definition can also give
 half-integer topological charges.
 (For SU($N_c$) gauge theory the topological charge can be an integer
 multiple of $1/N_c$.)
 Since $K^\prime$ is larger than the dynamical hopping parameter $K$
 and the level crossings typically occur between $K$ and $K^\prime$,
 it is plausible that the path integral at $K$ is dominated by
 configurations with positive Pfaffian.
 Therefore the Curci-Veneziano action gives the same continuum limit as
 the definition with the path integral over Majorana fermions.

 The question of fractional topological charges is important for the
 low energy effective action \cite{VENYAN} which assumes a remnant
 chiral symmetry $Z_{2N_c}$.
 In fact, if fractional topological charges would exist, the
 $\theta$-parameter would be periodic only with $2\pi N_c$ and the
 remnant chiral symmetry would be $Z_2$, instead of $Z_{2N_c}$.
 In this case the spontaneous symmetry breaking $Z_{2N_c} \to Z_2$ would
 be absent and there would be no first order phase transition at zero
 gluino mass.
 In the quenched continuum limit the half-integer topological charges
 seem to persist \cite{EHN_TOP} but the continuum limit with dynamical
 gluinos may be different.

\section{MULTI-BOSONIC ALGORITHM}\label{sec2}
 The multi-bosonic algorithm for Monte Carlo simulations of fermions has
 been proposed by L\"uscher \cite{LUSCHER}.
 In the original version for $N_f$ flavours one considers the
 approximation of the fermion determinant
\begin{eqnarray}\nonumber
\left|\det(Q)\right|^{N_f} 
& = & \left\{\det(Q^\dagger Q) \right\}^{N_f/2}
\\
\label{eq07}
& \simeq & \frac{1}{\det P_n(Q^\dagger Q)}
\end{eqnarray}
 where the polynomial $P_n$ satisfies
\be\label{eq08}
\lim_{n \to \infty} P_n(x) = x^{-N_f/2}
\ee
 in an interval $[\epsilon,\lambda]$ covering the spectrum of 
 $Q^\dagger Q$.
 For the multi-bosonic representation of the determinant one uses
\be\label{eq09}
P_n(Q^\dagger Q) = P_n(\tilde{Q}^2) = 
r_0 \prod_{j=1}^n (\tilde{Q}-\rho_j^*) (\tilde{Q}-\rho_j)
\ee
 and then
$$
 \prod_{j=1}^n\det[(\tilde{Q}-\rho_j^*) (\tilde{Q}-\rho_j)]^{-1} \propto
$$
\be\label{eq10}
\int [d\Phi] \exp\{ -\sum_{j=1}^n \sum_{xy}
\Phi_{jy}^+(\tilde{Q}-\rho_j^*) (\tilde{Q}-\rho_j)\Phi_{jx} \}
\ee
 The difficulty for small fermion masses is that the {\em condition
 number} $\lambda/\epsilon$ becomes very large ($10^4-10^6$) and very
 high orders $n = {\cal O}(10^3)$ are required for a good approximation.
 This requires large storage and the autocorrelation becomes bad since
 it is proportional to $n$.

\subsection{Improved version: two-step updating}\label{sec2.1}
 Substantially smaller storage and shorter autocorrelations for small
 fermion masses can be achieved by starting from a two-step
 approximation \cite{GLUINO}:
$$
\left\{\det(Q^\dagger Q) \right\}^{N_f/2}
$$
\be\label{eq11}
\;\simeq\;
\frac{1}{\det P^{(1)}_{n_1}(Q^\dagger Q) 
\det P^{(2)}_{n_2}(Q^\dagger Q)}
\ee
 where now
\be\label{eq12}
\lim_{n_2 \to \infty} P^{(1)}_{n_1}(x)P^{(2)}_{n_2}(x) = 
x^{-N_f/2} \ , \hspace{1em} 
x \in [\epsilon,\lambda] \ .
\ee
 The multi-bosonic representation of the determinant is used for
 $\det P^{(1)}_{n_1}$.
 The correction factor $\det P^{(2)}_{n_2}$ is realized in a {\em noisy
 correction step} \cite{KENKUT} with the accept-reject function
\be\label{eq13}
e^{-\eta^\dagger\left\{ P^{(2)}_{n_2}(\tilde{Q}[U^\prime]^2) -
                        P^{(2)}_{n_2}(\tilde{Q}[U]^2)\right\}\eta}
\ee
 where $\eta$ is generated from the simple Gaussian noise
 $\eta^\prime$ with a suitable polynomial approximation $P^{(3)}_{n_3}$
 as
\be\label{eq14}
\eta = P^{(2)}_{n_2}(\tilde{Q}[U]^2)^{-\half} \eta^\prime \Rightarrow
P^{(3)}_{n_3}(\tilde{Q}[U]^2) \eta^\prime \ .
\ee

 An important gain in performance is obtained by the use of
 quadratically (``least-square'') optimized polynomials \cite{POLYNOM}.
 The Chebyshev- (for $N_f=2$) or Legendre- (for $N_f=1$) polynomials are
 bad for large $\lambda/\epsilon$. 
 The quadratically optimized polynomials are much better.
 Note that, as discussed in sec.~\ref{sec1.1}, for gluinos one has to
 consider $N_f=\half$.

 A complete cycle of sweeps contains heatbath and overrelaxations sweeps
 for the boson fields and a Metropolis sweep for the gauge field 
 followed by the accept-reject step.
 We choose the order of the first polynomial such that the acceptance
 rate in the accept-reject step is typically 80-90\%.
 The longest autocorrelations appear for ``gluonic'' quantities
 depending on gauge field variables.
 For instance, in our runs the autocorrelation of the plaquette
 expectation value is typically given by
 $\tau_{int}^{plaq} \simeq 300-400$ cycles.
 The ``fermionic'' autocorrelations are about a factor $\simeq 10$
 smaller.

\subsection{Measurement correction}\label{sec2.2}
 To avoid systematic errors, both in the original version and in the
 two-step version the polynomial approximations have to be taken, in
 principle, to infinite order.
 In praxis it is enough that the errors introduced by the polynomial
 approximations are smaller than the statistical errors.
 However, due to the limited approximation quality, for small fermion
 masses sometimes {\em exceptional configurations} with very small
 ($10^{-8}-10^{-10}$) eigenvalues of $Q^\dagger Q$ are generated.
 In order to avoid these one could increase the polynomial orders but
 it is more economical to perform a {\em measurement correction} step
 by a generalization of the method of ref.~\cite{FREJAN}.
 In this way the expectation value of a quantity $A$ is given by
\be\label{eq15}
\langle A \rangle = \frac{
\langle A \exp{\{\eta^\dagger[1-P^{(4)}_{n_4}(Q^\dagger Q)]\eta\}}
\rangle_{U,\eta}}
{\langle  \exp{\{\eta^\dagger[1-P^{(4)}_{n_4}(Q^\dagger Q)]\eta\}}
\rangle_{U,\eta}} \ ,
\ee
 where $\eta$ is a Gaussian noise and the polynomial $P^{(4)}_{n_4}$ is
 such that $P^{(1)}_{n_1}P^{(2)}_{n_2}P^{(4)}_{n_4}$ is an optimized
 approximation of the function $x^{-N_f/2}$ in the interval
 $[\epsilon^\prime,\lambda]$.
 This correction step can be arranged in such a way that one obtains
 an ``exact'' algorithm.
 For instance, $\epsilon^\prime$ can be set to zero and then the
 correction step is not sensitive to the fluctuations of the smallest
 eigenvalue.
 (Note that in our case the least-square optimization can also be
 achieved for zero lower boundary of the interval, in contrast e.g.~to
 the minimization of the maximal relative deviation with Chebyshev
 polynomials.)
 For the evaluation of $P^{(4)}_{n_4}$ one can use $n_4$-independent
 recursive relations \cite{POLYNOM} which can be stopped by observing
 the convergence of the result.
 Another possibility is to take $\epsilon^\prime>0$ and to perform the
 correction exactly for the few exceptional eigenvalues below
 $\epsilon^\prime$ by multiplying with $\lambda^{N_f/2}
 P^{(1)}_{n_1}(\lambda)P^{(2)}_{n_2}(\lambda)$.

\begin{figure}[th]
\vspace*{-1.0cm}
\begin{center}
\epsfig{file=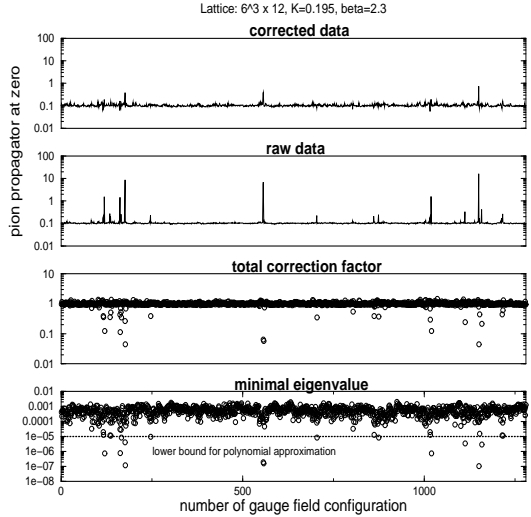,
        width=7.0cm,height=7.0cm,
        angle=0}
\vspace*{-1.0cm}
\caption{\label{fig03}
 The measurement correction for the a-pion propagator at zero distance.
 The exceptional configurations with small eigenvalues contribute
 strongly to the raw data.
 After correction these contributions are still important but of 
 normal size.}
\end{center}
\vspace*{-1.0cm}
\end{figure}

 Experience shows that for most physical quantities the exceptional
 configurations give no exceptional contributions.
 There are, however, also some quantities which are sensitive to the
 small eigenvalues, hence the measurement correction is essential
 (see fig.~\ref{fig03}).
 As the figure shows, even the small lower limit of the polynomial
 approximation interval $\epsilon=10^{-5}$, in connection with the
 polynomial orders $n_1=24,\; n_2=200$, is insufficient for suppressing
 the exceptional configurations.
 In this case a measurement correction step with $\epsilon^\prime=0$ and
 $n_4=400$ is enough for good precision.

\section{APPROACHING \\ SUPERSYMMETRY}\label{sec3}
\begin{figure}[th]
\vspace*{-1.2cm}
\begin{center}
\epsfig{file=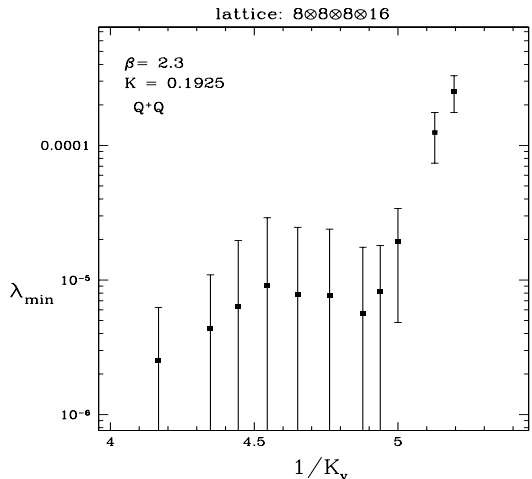,
        width=7.5cm,height=6.4cm,
        angle=0}
\vspace*{-1.5cm}
\caption{\label{fig04}
 The average lowest eigenvalue of $Q^\dagger Q$ as a function of the
 inverse (valence) hopping parameter in a dynamical update at
 ($\beta=2.3,K=0.1925$) on $8^3 \cdot 16$ lattice.
 The error bars here show standard deviations (not the errors of
 averages).}
\end{center}
\vspace*{-1.0cm}
\end{figure}

 The Monte Carlo simulations have been performed up to now mainly at
 $\beta=2.3$.
 After last years tests at lower hopping parameter values
 \cite{DESYMUNSTER} we are presently concentrating on the interval
 $0.185 \leq K \leq 0.1975$.
 Lattice sizes range from $4^3 \cdot 8$ to $12^3 \cdot 24$.
 The characteristic feature of these simulations are the rather large
 condition numbers growing up to $\lambda/\epsilon = 10^5-10^6$.
 Our present best estimate for the critical hopping parameter for zero
 gluino mass is $K_0 \simeq 0.195$ (see sec.~\ref{sec3.3}).
 This means that the bare gluino mass in lattice units at
 e.g.~$K=0.1925$ is about
 $am_0 \equiv \half(K_0^{-1}-K^{-1}) \simeq 0.03$.
 This gives a ratio to the smallest glueball mass
 $m_0/M^{0^+} \simeq 0.08$.
 The large condition numbers are further increased if the ``valence''
 hopping parameter $K_v$ in the gluino propagators is increased.
 A typical example for the average of the lowest eigenvalue is shown in
 fig.~\ref{fig04}.
 (The approximation interval for polynomials is in this case
 $[\epsilon,\lambda]=[0.0001,3.7]$.)
 As one can see, near $1/K_v=1/0.2025 \simeq 4.94$ the average smallest
 eigenvalue becomes rather small and fluctuates practically down to
 zero.
 For smaller values of $1/K_v$ the picture remains unchanged.
 This agrees with the findings of ref.~\cite{EHN_FLOW} about the absence
 of the spectrum gap for a range of bare fermion masses.
 (Note that the start of the gapless region at $K_v=0.2025$ lies higher
 than $K_0$ because in fig.~\ref{fig04} the dynamical ``sea'' gluino
 mass is not changing.)

\subsection{Bound state masses}\label{sec3.1}
 A basic assumption about the dynamics of SYM theory is that there is 
 confinement.
 The colourless bound states should be organized in degenerate
 supermultiplets which are split up at non-zero gluino mass.
 The pattern of states should be similar to QCD with
 a single flavour of quarks.
 For instance, the low-lying meson states have quantum numbers as
 $\eta^\prime,\; f_0$ etc.
 Since our constituent gluinos are in the adjoint representation, we
 call these states $a$-$\eta^\prime,\; a$-$f_0$ etc.
 There are, of course, also the glueball states ($gg$) as in pure gauge
 theory and for completing the supermultiplets we also need spin-$\half$
 states: the {\em gluino-glueballs} ($\tilde{g}g$).

\begin{figure}[th]
\vspace*{-1.0cm}
\begin{center}
\epsfig{file=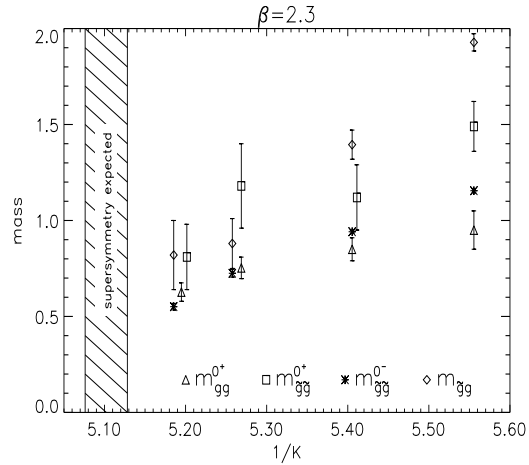,
        width=7.5cm,height=6.5cm,
        angle=0}
\vspace*{-1.5cm}
\caption{\label{fig05}
 The preliminary results for bound state masses.}
\end{center}
\vspace*{-1.0cm}
\end{figure}

 We determine the masses of low-lying states by using the
 blocking-smearing technology for achieving better projections to the
 sources and sinks.
 Our preliminary results are summarized in fig.~\ref{fig05}.
 The presence of two distinct light $0^+$ states in the spectrum seems
 to prefer the generalization of the Veneziano-Yankielowicz action
 \cite{VENYAN} proposed in ref.~\cite{FAGASCH}.

\subsection{String tension}\label{sec3.2}
\begin{figure}[th]
\vspace*{-1.0cm}
\begin{center}
\epsfig{file=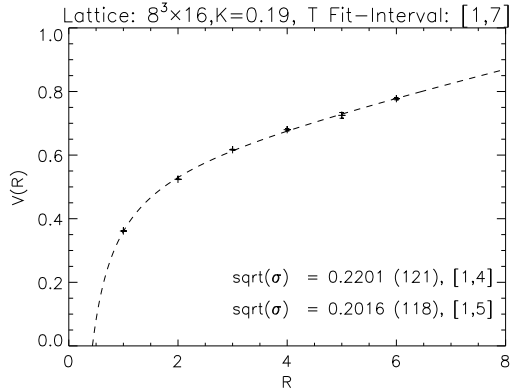,
        width=7.5cm,height=5.5cm,
        angle=0}
\vspace*{-1.5cm}
\caption{\label{fig06}
 The three-parameter fit to the static potential:
 $V(R)=V_0+\sigma R-\alpha/R$ at ($\beta=2.3,K=0.19$) on $8^3 \cdot 16$
 lattice.}
\end{center}
\vspace*{-1.2cm}
\end{figure}
 The confinement can be characterized by the string tension obtained
 from the linear part of the static potential at large distances.
 We determined the static potential from APE-smeared Wilson loops
 following \cite{WUPPERTAL}.
 At ($\beta=2.3,K=0.19$) on $8^3 \cdot 16$ lattice the optimized
 smearing radius is 3.3 and we obtained stable results against
 variations of the $T$ fit-interval.
 As shown by fig.~\ref{fig06}, the result for the square root of the
 string tension in lattice units is $a\sqrt{\sigma}=0.21(1)$.
 At $K=0.1925$ the corresponding value is surprisingly small:
 $a\sqrt{\sigma}=0.10(3)$ but the $8^3 \cdot 16$ lattice might be too
 small in this case, causing finite volume effects.

\subsection{Gluino condensate}\label{sec3.3}
\begin{figure}[th]
\vspace*{-1.2cm}
\begin{center}
\epsfig{file=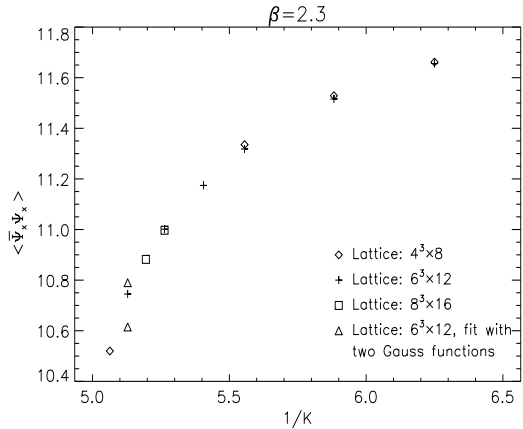,
        width=7.5cm,height=6.0cm,
        angle=0}
\vspace*{-1.5cm}
\caption{\label{fig07}
 The preliminary results for the gluino condensate.}
\end{center}
\vspace*{-1.0cm}
\end{figure}

 If there are no fractional topological charges, the hopping parameter
 $K_0$ corresponding to zero gluino mass is signaled by a first order
 phase transition which is due to the spontaneous discrete chiral
 symmetry breaking $Z_4 \to Z_2$.
 The gluino condensate should have a jump of order ${\cal O}(1)$ in
 lattice units \cite{BUCKOW}. 
 It is possible that this transition develops only in the continuum 
 limit and at finite lattice spacings it shows up just as a fast
 cross-over.
 Our preliminary results are summarized in fig.~\ref{fig07}.
 The jump at $1/K=1/0.195 \simeq 5.13$ suggests a vanishing gluino mass
 at $K_0 \simeq 0.195$.

 {\bf Acknowledgements:} It is a pleasure to thank Gernot M\"unster
 for helpful discussions.
 J. Westphalen acknowledges the financial contribution of the European
 Commission under the TMR-Program ERBFMRX-CT97-0122.


\end{document}